# Estimators in Cryptography

**Ph.D. Nicolae Constantinescu**
**Faculty of Mathematics and Informatics, Department of Informatics,**
**University of Craiova**
nikyc@central.ucv.ro

**ABSTRACT.** One of the main problems in cryptography is to give criteria to provide good comparators of cipher systems. The security of a cipher system must include the security of the algorithm, the security of the key generator and management module (see [BM94], [CM97],[Mau92a]) and the security of the cryptographic key agreement protocol (see [Mau93a],[MC94],[Mau93b],[Mau92b]).
This paper gives show the necessary mathematical background to estimate the most important cryptographic measures of the key generators and of the unconditionally key agreement protocols. These cryptographic measures are the Shannon entropy (for the key generator module) and Renyi entropy of order $\alpha$ for the key agreement protocol.

## 1   Introduction

It is known that Shannon entropy is a limit case ($\alpha \rightarrow 1$) for the Renyi entropy. From this reason we focus on an estimating method of Renyi entropy. This method will be the Bayesian one (see [Pre91], [Gui77]) which is based on mixing priori information regarding $\theta \in \Theta$ given by the distribution $\pi(\theta)$, with the selection information given by X after the experimentation process. We denote with $\pi(\theta|x)$ the posteriori distribution of $\theta$ conditioned by X and with $f(x|\theta)$ the density of probability of the random variable X in the hypotheses that the true state of the parameter is $\theta$.

The density of X is given by

$$f(x) = E^{\pi}(f(x \mid \theta) = \int_{\Theta} f(x \mid \theta)\pi(\theta)d\theta$$

Using Bayes formula we get and the posteriori distribution

In general f(x) and $\pi(\theta|x)$ are hard to compute and from this reason we must estimate it using computational techniques such as Monte Carlo

$$\pi(\theta \mid x) = \frac{\pi(\theta)f(x \mid \theta)}{\int_{\Theta} \pi(\theta)f(x \mid \theta)d\theta}$$

methods.

This paper compute Bayesian estimators for the Renyi entropy of order $\alpha$ and we proof that the Bayesian estimator of Shannon entropy is obtained as an limiting case $\alpha \rightarrow 1$ and we use like distribution on the frequency space the multinomial distribution and like a priori the Dirichlet distribution because this Distribution is conjugate for the multinomial distribution family. We also present the Bayesian - Monte Carlo estimator. Our work is a generalization of an recently result obtained by [YK97]. In general an information source (see [Gui77]) can be approximated by its L-order approximation (the cardinal of the symbol space of the source output is L) and we study the asymptotic behavior of the estimators and we proof that the approximation method is asymptotic regarding the codification source parameter. These asymptotic results are generalizations of the results obtained by Maurer in [Mau97] and can be used to estimate the effective size of a cipher system.

## 2 Estimators

The fundamental problem in cryptography is the generation of a shared secret key by two parties, A and B, not sharing a secret key initially over an insecure channel which is under the control of E. The general mathematical model proposed in [BM94] is that in which A and B are connected only by a public channel and E can eavesdrop the communication. The problem can be solved with public key cryptography in which we assume that the power of computing of E is limited. Another possibility is to develop techniques that avoid the above assumption. The motivation for is two-fold: First, one avoids having to worry about the generality of a particular computational model, which is of some concern in view of the potential feasibility of quantum computers (see [BB92]). Secondly, and more importantly, no strong rigorous results on the difficulty of breaking a cryptosystem have been proved, and this problem continues to be among the most difficult ones in complexity theory.

## 2.1   Cryptographic preliminaries

The general protocol take place in a scenario where A, B and E know the correlated random variables X, Y, Z, respectively, distributed according to some joint probability distribution that may be under partial control of E (like for the case of quantum cryptography).
We can see that the problem can be solved in the following phases:
   1. A and B must detect any modification or insertion of messages.
   2. A and B establish a secret communication key.

   The first phase is called authentication step. This can be done with classical statistical tests (see [Bla87] and [Mau97]).
   The second phase consists in three steps:
   a) *Advantage distillation* The purpose of this step is to create a random variable about either A or B has more information than E. Advantage distillation is only needed when W is not immediately available from X and Y. A and B create W by exchanging messages, summarized as the random variable C, over a public channel. A discussion on these facts can be found in [Mau93b].
   b) *Information reconciliation.* To agree on a string T with very high probability, A and B exchange redundant error-correction information U, such as a sequence of parity checks. After this phase, E (incomplete) information about T consists of Z, C and U (see [Mau93a]).
   c) *Privacy amplification.* In the final phase, A and B agree publicly on a compression function G to distill a shorter string S about which E has only a negligible amount of information (see [BM94] and [Cac97]). Therefore, S can be subsequently be used as a secret key. In [Cac97] and [CM97] Cachin proof the connection between smooth entropy, Renyi entropy and privacy amplification phase. In this paper we study the effect of side information U on the collision entropy (Renyi entropy of order 2) which is a measure of the security of the protocol.

## 2.2 Mathematical overwiew

We assume that the reader is familiar with the notion of entropy and the basic concepts of information theory (see [Bla87]). In privacy amplification, a different and a non-standard entropy measure, collision entropy, is of central importance. Collision entropy is also known as Renyi entropy of order 2 (see [Cac97]).

**Definition 1** (collision probability -see [Mau92a]). *Let X be a random variable with the alphabet* X *and the distribution* $P_X$. *The Collision probability* $P_C(X)$ *of the random variable X is defined as the probability that X takes on the same value twice in two independent experiments, i.e.,*

$$P_C(X) = \sum_{x \in X} P_X(x)^2$$

**Definition 2** (collision entropy -see [Mau92a]). *Collision entropy of the random variable X is* $H_C(X) = -\log P_C(X)$

**Remark**. We see that collision entropy is Renyi entropy of order $\alpha = 2$ which is defined like $H_\alpha(X) = \frac{1}{1-\alpha} \log \sum_{x \in X} P_X^\alpha(x)$. In the limit case $\alpha \to 1$ we get $H(X) = \lim_{\alpha \to 1} H_\alpha(X)$ (Shannon entropy) and when $\alpha \to \infty$ we obtain the *min-entropy* $H_\infty(X) = -\log \max_{x \in X} P(x)$. We also have the following inequalities $\log|X| \geq H(X) \geq H_2(X) \geq H_\infty(X)$ and $0 < \alpha < \beta \Rightarrow H_\infty(X) \geq H_\beta(X)$ with equality iff X is uniformly distributed over X or a subset of X.

**Definition 3** (conditioned collision entropy see [Mau92a]). *For a event* $\varepsilon$ *the collision entropy of X conditioned on varepsilon,* $H\_c(X|\varepsilon)$ *is defined naturally as the collision entropy of the conditional distribution* $P\_X|\varepsilon$. *The collision entropy conditioned on a random variable,* $H\_c(X|Y)$ *is defined as the expected value of the conditional collision entropy:*

$$H_C(X|Y) = \sum_y P_Y(y) H_C(x|Y=y) = E^{P_Y}[H_C(X|Y)]$$

In [MPV97] and [MMP97] Morales presented same of the applications of $\phi$-entropies to comparison of experiments. A similar work on comparison of experiments is done in [PM93],[PMT91],[PM+97], [PMP93] by Pardo who considered generalized entropy measures. Therefore is very interesting to study the Bayesian estimation of such entropies which includes Shannon entropy and collision entropy. The sample behavior of entropy parameters is done in [PM93].

If we denote $\alpha_i = \alpha\pi_i$ , with $\sum_{i=1}^{s}\pi_i = 1$ by straightforward calculations we have for s $\geq$ 2 :

$$E_n^\gamma = -\sum_{i=1}^{s}\frac{\Gamma(\alpha+n)}{\Gamma(\alpha\pi_i+n_i)}\frac{\Gamma(\alpha\pi_i+n+\gamma)}{\Gamma(\alpha+\gamma+n)}$$

**Remarks.**

a) The estimation of $H_\gamma(X) = \frac{1}{1-\gamma}\log\sum_{x\in X}P_X^\gamma(x)$ will be $H_{n,\gamma}^B(X) = \frac{1}{1-\gamma}\log E_n^\gamma$ for $\gamma \to 1$ we obtain after L'Hospital rule the results obtained by *Yuan* in [YK97].

b) For $\gamma = 2$ we obtain the Bayesian estimation of collision entropy which is a measure of protocol security in key agreement protocols over an insecure channel.

The implementation of this mathematical approach has been started to become used in applications since the experimental quantum computing (see [BB92]), in Cryptanalysis (see [SC00]) and in Reliability Estimations (see [Sha93]).